\documentclass[prb,showpacs,twocolumn]{revtex4}%
\usepackage{graphicx}
\usepackage{dcolumn}
\usepackage{bm}
\usepackage{amsmath}
\usepackage{amsfonts}
\usepackage{amssymb}%
\providecommand{\U}[1]{\protect\rule{.1in}{.1in}}
\begin{document}
\title{The effects of k-dependent self-energy \\in the electronic structure of
correlated materials}
\author{T. Miyake$^{1,3}$, C. Martins$^{1,3}$, R. Sakuma$^{2}$, and F.
Aryasetiawan$^{2,3}$}
\affiliation{$^{1}$Nanosystem Research Institute ``RICS'', AIST, Tsukuba 305-8568, Japan}
\affiliation{$^{2}$Depatment of Physics, Division Mathematical Physics, Lund University,
S\"{o}lvegatan 14A, 223 62 Lund, Sweden,}
\affiliation{$^{3}$Japan Science and Technology Agency, CREST, Kawaguchi, Saitama 332-0012, Japan}

\begin{abstract}
It is known from self-energy calculations in the electron gas and $sp$
materials based on the $GW$ approximation that a typical quasiparticle
renormalization factor ($Z$ factor) is approximately $0.7-0.8$. Band narrowing
in electron gas at $r_{s}=4$ due to correlation effects, however, is only
approximately $10\%$, significantly smaller than the $Z$ factor would suggest.
The band narrowing is determined by the frequency-dependent self-energy,
giving the $Z$ factor, and the momentum-dependent or nonlocal self-energy. The
results for the electron gas point to a strong cancellation between the
effects of frequency- and momentum-dependent self-energy. It is often assumed
that for systems with a narrow band the self-energy is local. In this work we
show that even for narrow-band materials, such as SrVO$_{3}$, the nonlocal
self-energy is important.

\end{abstract}

\maketitle

\section{Introduction}

The electronic structure of a certain class of materials, commonly known as
strongly correlated materials, is crucially determined by electron
correlations. In these materials one-particle theory, predominantly the local
density approximation (LDA) within density functional theory\cite{Kohn}, is
often far from sufficient in providing an accurate and reliable description of
the electronic structure. The reason for the failure of one-particle theory
may be traced back to the presence of many configurations close in energy
arising from a partially filled narrow band characteristic of these materials,
similar to the situation in atoms with partially filled shell. It is not
surprising that description of the electronic structure in terms of a single
Slater determinant is not satisfactory. One of the most fruitful approaches in
treating the electronic structure of these materials is the Green function
technique with a nonlocal and energy-dependent self-energy. Among the Green
function approaches, the combination of the LDA and the dynamical mean-field
theory (DMFT)\cite{Georges96}, i.e., the LDA+DMFT
method\cite{Anisimov97,Lichtenstein98}, is perhaps the most widely used
technique in describing the electronic structure of strongly correlated materials.

In many calculations based on model Hamiltonians it is often assumed that the
self-energy is local, meaning that it has no momentum or $\mathbf{k}$
dependence and the Hubbard $U$ used in solving the impurity problem is assumed
to be static. How these two approximations affect the resulting band structure
remains to be investigated. Experience with the electron gas teaches us that
there is a strong cancellation between the effects of $\mathbf{k}$-dependent
self-energy and frequency-dependent self-energy. This is revealed by the fact
that effective mass of the electron gas with density relevant for most
materials is quite close to unity. The effective mass in general depends on
two factors: the derivative of the self-energy with respect to frequency,
which tends to enhance the effective mass, and the derivative of the
self-energy with respect with the $\mathbf{k}$-vector, which tends to have the
opposite effect. In the electron gas these two factors tend to cancel each
other, hence the effective mass being close to unity\cite{vonBarth96,Holm98}.
We have no apriori reason to believe that such an almost complete cancellation
remains true in real materials, especially in the strongly correlated systems.
Indeed, it is often claimed that for narrow-band materials the self-energy has
little $\mathbf{k}$-dependence so that a local but dynamic theory with
frequency-dependent self-energy provides a good approximation to the full self-energy.

The $\mathbf{k}$-dependent self-energy becomes even more relevant with the
recent progress in solving the impurity problem with a frequency-dependent
Hubbard $U$\cite{Werner07,Werner10,Werner06}. This new algorithm makes it
possible to perform LDA+DMFT calculations using a dynamic Hubbard \emph{U},
instead of a static one as in conventional calculations. This raises an
important issue concerning the role of nonlocal or $\mathbf{k}$-dependent
self-energy. The use of dynamic \emph{U} might lead to an overestimation of
band narrowing or mass enhancement because the dynamic \emph{U} produces a
self-energy that has a stronger frequency dependence or larger derivative at
around the Fermi level. One might anticipate qualitatively that the use of a
dynamic \emph{U} ought to be counterbalanced by the inclusion of a
$\mathbf{k}$-dependent self-energy to restore the correct mass enhancement.

The purpose of this paper is to investigate the effects of momentum- and
frequency-dependence of the self-energy on the electronic structure of
materials with partially filled narrow bands. To address the forementioned
issues, instead of using a model Hamiltonian, we will employ the \emph{GW}
approximation (GWA)\cite{Hedin65,ferdi98} which allows us to perform accurate
calculations on real materials. As a test material we study specifically the
electronic structure of SrVO$_{3}$ which has been widely studied in the
literature both experimentally by means of photoemission
spectroscopy\cite{morikawa95,inoue98,sekiyama04,yoshida05,takizawa09,yoshida10,yoshimatsu11,aizaki12}
and theoretically\cite{pavarini04,nekrasov06,karolak11,casula12}. We believe
the results are general and will be directly relevant to realistic electronic
structure calculations.

\section{Theory and Method}

The correlation part of the self-energy (excluding exchange) in the GWA is
given by\cite{Hedin65,ferdi98}%

\begin{equation}
\Sigma^{c}(\mathbf{r,r}^{\prime};\omega)=i\int\frac{d\omega^{\prime}}{2\pi
}G(\mathbf{r,r}^{\prime};\omega+\omega^{\prime})W^{c}(\mathbf{r,r}^{\prime
};\omega^{\prime})
\end{equation}
where%

\begin{equation}
W^{c}=W-v.
\end{equation}
$v$ is the bare Coulomb interaction and $W$ is the fully screened interaction
calculated within the random-phase approximation (RPA). Using the spectral
representation of $G$ and $W$%

\begin{equation}
G(\mathbf{r,r}^{\prime};\omega)=\sum_{\mathbf{k}n}^{\text{occ}}\frac
{\psi_{\mathbf{k}n}(\mathbf{r)}\psi_{\mathbf{k}n}^{\ast}(\mathbf{r}^{\prime}%
)}{\omega-\varepsilon_{\mathbf{k}n}-i\delta}+\sum_{\mathbf{k}n}^{\text{unocc}%
}\frac{\psi_{\mathbf{k}n}(\mathbf{r)}\psi_{\mathbf{k}n}^{\ast}(\mathbf{r}%
^{\prime})}{\omega-\varepsilon_{\mathbf{k}n}+i\delta},
\end{equation}

\begin{equation}
W^{c}(\mathbf{r,r}^{\prime};\omega)=\int_{-\infty}^{0}d\omega^{\prime}%
\frac{B(\mathbf{r,r}^{\prime};\omega^{\prime})}{\omega-\omega^{\prime}%
-i\delta}+\int_{0}^{\infty}d\omega^{\prime}\frac{B(\mathbf{r,r}^{\prime
};\omega^{\prime})}{\omega-\omega^{\prime}+i\delta},
\end{equation}
where%

\begin{equation}
B(\omega)=-\frac{1}{\pi}\operatorname{Im}W^{c}(\omega)\text{sgn}(\omega),
\end{equation}
the spectral function of the correlation part of the self-energy can be
expressed in terms of the imaginary part of the screened interaction as
follows\cite{ferdi92}:%

\begin{equation}
\Gamma(\mathbf{r,r}^{\prime};\omega\leq\mu)=\sum_{\mathbf{k}n}^{\text{occ}%
}\psi_{\mathbf{k}n}(\mathbf{r)}B(\mathbf{r,r}^{\prime};\varepsilon
_{\mathbf{k}n}-\omega)\psi_{\mathbf{k}n}^{\ast}(\mathbf{r}^{\prime}%
)\theta(\varepsilon_{\mathbf{k}n}-\omega), \label{Gamma1}%
\end{equation}

\begin{equation}
\Gamma(\mathbf{r,r}^{\prime};\omega>\mu)=\sum_{\mathbf{k}n}^{\text{occ}}%
\psi_{\mathbf{k}n}(\mathbf{r)}B(\mathbf{r,r}^{\prime};\omega-\varepsilon
_{\mathbf{k}n})\psi_{\mathbf{k}n}^{\ast}(\mathbf{r}^{\prime})\theta
(\omega-\varepsilon_{\mathbf{k}n}), \label{Gamma2}%
\end{equation}
where%

\begin{equation}
\Gamma(\mathbf{r,r}^{\prime};\omega)=-\frac{1}{\pi}\operatorname{Im}\Sigma
^{c}(\mathbf{r,r}^{\prime};\omega)\text{sgn}(\omega-\mu).
\end{equation}
The real part of the correlation self-energy (excluding exchange) is given by
the Hilbert transform%

\begin{equation}
\operatorname{Re}\Sigma^{c}(\mathbf{r,r}^{\prime};\omega)=\int_{-\infty
}^{\infty}d\omega^{\prime}\frac{\Gamma(\mathbf{r,r}^{\prime};\omega^{\prime}%
)}{\omega-\omega^{\prime}}.
\end{equation}
In this work, the one-particle band structure $\left\{  \psi_{\mathbf{k}%
n},\varepsilon_{\mathbf{k}n}\right\}  $ is taken to be the LDA one.

The full self-energy may be expanded in terms of Bloch states $\psi
_{\mathbf{k}n}$:%

\begin{equation}
\Sigma(\mathbf{r,r}^{\prime};\omega)=\sum_{\mathbf{k}nn^{\prime}}%
\psi_{\mathbf{k}n}(\mathbf{r)\Sigma}_{nn^{\prime}}(\mathbf{k},\omega
)\psi_{\mathbf{k}n}^{\ast}(\mathbf{r}^{\prime}).
\end{equation}
The Bloch states may be expressed in terms of some Wannier orbitals $\left\{
\varphi_{\mathbf{R}n}\right\}  $%

\begin{equation}
\psi_{\mathbf{k}n}(\mathbf{r)=}\sum_{\mathbf{R}}\exp(-i\mathbf{k\cdot
R)}\varphi_{\mathbf{R}n}(\mathbf{r})
\end{equation}
and the self-energy becomes%

\begin{align}
&  \Sigma(\mathbf{r,r}^{\prime};\omega)\nonumber\\
&  =\sum_{\mathbf{k}nn^{\prime}}\sum_{\mathbf{RR}^{\prime}}\exp
[-i\mathbf{k\cdot(R-R}^{\prime})\mathbf{]}\varphi_{\mathbf{R}n}(\mathbf{r}%
)\mathbf{\Sigma}_{nn^{\prime}}(\mathbf{k},\omega)\varphi_{\mathbf{R}^{\prime
}n^{\prime}}^{\ast}(\mathbf{r}^{\prime}).
\end{align}
The local self-energy centered at lattice site $\mathbf{R}$ is defined
according to%

\begin{align}
\Sigma_{\mathbf{R}}^{\text{loc}}(\mathbf{r,r}^{\prime};\omega)  &
=\sum_{\mathbf{k}nn^{\prime}}\varphi_{\mathbf{R}n}(\mathbf{r})\mathbf{\Sigma
}_{nn^{\prime}}(\mathbf{k},\omega)\varphi_{\mathbf{R}n^{\prime}}^{\ast
}(\mathbf{r}^{\prime})\nonumber\\
&  =\sum_{nn^{\prime}}\varphi_{\mathbf{R}n}(\mathbf{r})\mathbf{\Sigma
}_{nn^{\prime}}^{\text{loc}}(\omega)\varphi_{\mathbf{R}n^{\prime}}^{\ast
}(\mathbf{r}^{\prime}),
\end{align}
where%

\begin{equation}
\mathbf{\Sigma}_{nn^{\prime}}^{\text{loc}}(\omega)=\sum_{\mathbf{k}%
}\mathbf{\Sigma}_{nn^{\prime}}(\mathbf{k},\omega). \label{Sigmalocal}%
\end{equation}
Since $\Sigma_{\mathbf{R}}^{\text{loc}}$ is independent of $\mathbf{R}$ we may
choose $\mathbf{R}=0$ and write%

\begin{equation}
\Sigma^{\text{loc}}(\mathbf{r,r}^{\prime};\omega)=\sum_{nn^{\prime}}%
\varphi_{n}(\mathbf{r})\mathbf{\Sigma}_{nn^{\prime}}^{\text{loc}}%
(\omega)\varphi_{n^{\prime}}^{\ast}(\mathbf{r}^{\prime}) \label{Sigmaloc}%
\end{equation}
where it is understood that%

\begin{equation}
\varphi_{n}\doteqdot\varphi_{\mathbf{0}n}.
\end{equation}

The above formulation is quite general and we now focus on SrVO$_{3}$ which
has a narrow band, derived from the $t_{2g}$ orbitals of the vanadium,
crossing the Fermi level. The $t_{2g}$ band is well separated from the rest of
the bands so that we need only consider Wannier orbitals constructed from
Bloch states belonging to this band. Following the method of Marzari and
Vanderbilt\cite{Marzari97,Souza01}, we construct the (maximally localized)
Wannier orbitals $\varphi_{\mathbf{R}n}$ according to%

\begin{equation}
\left\vert \varphi_{m}^{\mathbf{R}}\right\rangle =\sum_{\mathbf{k}}\left\vert
\psi_{\mathbf{k}m}^{w}\right\rangle e^{i\mathbf{k\cdot R}},\ \ \left\vert
\psi_{\mathbf{k}m}^{w}\right\rangle =\frac{1}{\sqrt{N}}\sum_{\mathbf{R}%
}\left\vert \varphi_{m}^{\mathbf{R}}\right\rangle e^{-i\mathbf{k\cdot R}}
\label{Wannier}%
\end{equation}

\begin{equation}
\left\vert \psi_{\mathbf{k}m}^{w}\right\rangle =\sum_{n}\left\vert
\psi_{\mathbf{k}n}\right\rangle S_{nm}(\mathbf{k}),\ m,n\in t_{2g}
\label{Wannier-gauge}%
\end{equation}
$\left\vert \psi_{\mathbf{k}n}\right\rangle \ $is a one-particle Bloch state
which may be chosen to be that of the LDA. The matrix $S$ is a square matrix
when the subspace is well separated but for entangled bands, $S$ is not
necessarily a square matrix, the number of band index $n$ may be larger than
$m$. The matrix $S$ is chosen to make the Wannier orbitals as localized as
possible according to a prescription by Marzari and
Vanderbilt\cite{Marzari97,Souza01}.

Using the expression in (\ref{Sigmaloc}) the expectation value of the local
self-energy in a Bloch state $\psi_{\mathbf{k}n}$ is given by%

\begin{equation}
\left\langle \psi_{\mathbf{k}n}|\Sigma^{\text{loc}}(\omega)\psi_{\mathbf{k}%
n}\right\rangle =\sum_{mm^{\prime}\subset t_{2g}}\left\langle \psi
_{\mathbf{k}n}|\varphi_{m}\right\rangle \Sigma_{mm^{\prime}}^{\text{loc}%
}(\omega)\left\langle \varphi_{m^{\prime}}|\psi_{\mathbf{k}n}\right\rangle
\label{Siglocnn}%
\end{equation}
where $\varphi_{m}$ is a maximally localized Wannier orbital defined in
(\ref{Wannier}) for the central cell $\mathbf{R}=0$ and the matrix elements
$\Sigma_{mm^{\prime}}^{\text{loc}}$ are taken in the Wannier gauge
$\psi_{\mathbf{k}m}^{w}$ defined in Eq. (\ref{Wannier-gauge}):%

\begin{equation}
\Sigma_{mm^{\prime}}^{\text{loc}}(\omega)=\sum_{\mathbf{k}}\left\langle
\psi_{\mathbf{k}m}^{w}|\Sigma^{\text{loc}}(\omega)|\psi_{\mathbf{k}m^{\prime}%
}^{w}\right\rangle . \label{Siglocmm}%
\end{equation}
The sum in Eq. (\ref{Siglocnn}) is restricted to the $t_{2g}$ orbitals because
the Bloch state $\psi_{\mathbf{k}n}$ belonging to the $t_{2g}$ band has no
component outside the $t_{2g}$ subspace. In the case of SrVO$_{3}$, due to
cubic symmetry, the self-energy in Eq. (\ref{Siglocmm}) is diagonal and
independent of the orbital index $m$ so that the matrix element of the local
self-energy in the Bloch state $\psi_{\mathbf{k}n}$ in Eq. (\ref{Siglocnn})
becomes independent of both $\mathbf{k}$ and the band index $n$.

The quasiparticle band structure is obtained from%

\begin{align}
E_{\mathbf{k}n}  &  =\varepsilon_{\mathbf{k}n}+\left\langle \psi_{\mathbf{k}%
n}|\operatorname{Re}\Sigma(E_{\mathbf{k}n})-v_{xc}|\psi_{\mathbf{k}%
n}\right\rangle \nonumber\\
&  \approx\varepsilon_{\mathbf{k}n}+Z_{\mathbf{k}n}\left\langle \psi
_{\mathbf{k}n}|\operatorname{Re}\Sigma(\varepsilon_{\mathbf{k}n})-v_{xc}%
|\psi_{\mathbf{k}n}\right\rangle ,
\end{align}
where%

\begin{equation}
Z_{\mathbf{k}n}=\left[  1-\frac{\partial\operatorname{Re}\Sigma_{nn}%
(\mathbf{k},\omega)}{\partial\omega}\right]  _{\omega=\varepsilon
_{\mathbf{k}n}}^{-1}%
\end{equation}
is the quasiparticle weight. The angle-resolved spectral function is
calculated as follows:%

\begin{equation}
A(\mathbf{k},\omega)=\frac{1}{\pi}\sum_{n}\frac{|\operatorname{Im}\Sigma
_{nn}(\mathbf{k},\omega)|}{\left[  \omega-\varepsilon_{\mathbf{k}%
n}-\operatorname{Re}\Sigma_{nn}(\mathbf{k},\omega)\right]  ^{2}+\left[
\operatorname{Im}\Sigma_{nn}(\mathbf{k},\omega)\right]  },
\end{equation}
where we have assumed that the self-energy is diagonal in the band index, and
the total spectral function is given by%

\begin{equation}
A(\omega)=\sum_{\mathbf{k}}A(\mathbf{k},\omega).
\end{equation}

The band structure calculation is based on the full-potential LMTO
implementation.\cite{methfessel00} The exchange-correlation functional is the
local density approximation of the Cepeley-Alder type~\cite{ceperley80}. The
GW calculation uses mixed basis consisting of products of two atomic orbitals
and interstitial plane waves.\cite{ferdi94,schilfgaarde06} The 8$\times
$8$\times$8 mesh is used for Brillouin-zone integration. More technical
details are found elsewhere.\cite{miyake08}

\section{Results and Discussions}

\subsection{Local vs nonlocal self-energy}

\begin{figure}[ptb]
\begin{center}
\includegraphics[width=85mm]{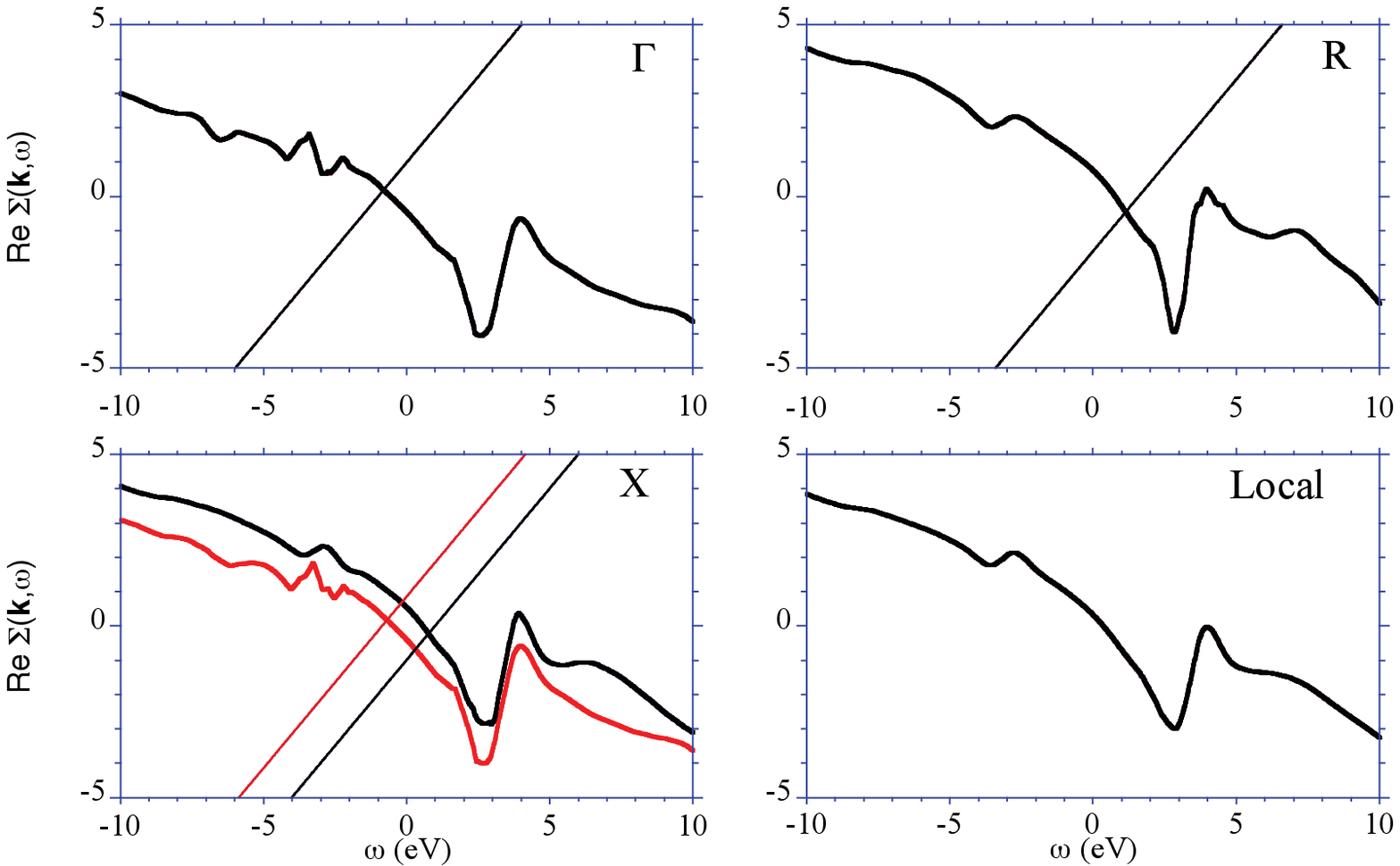} 
\includegraphics[width=85mm]{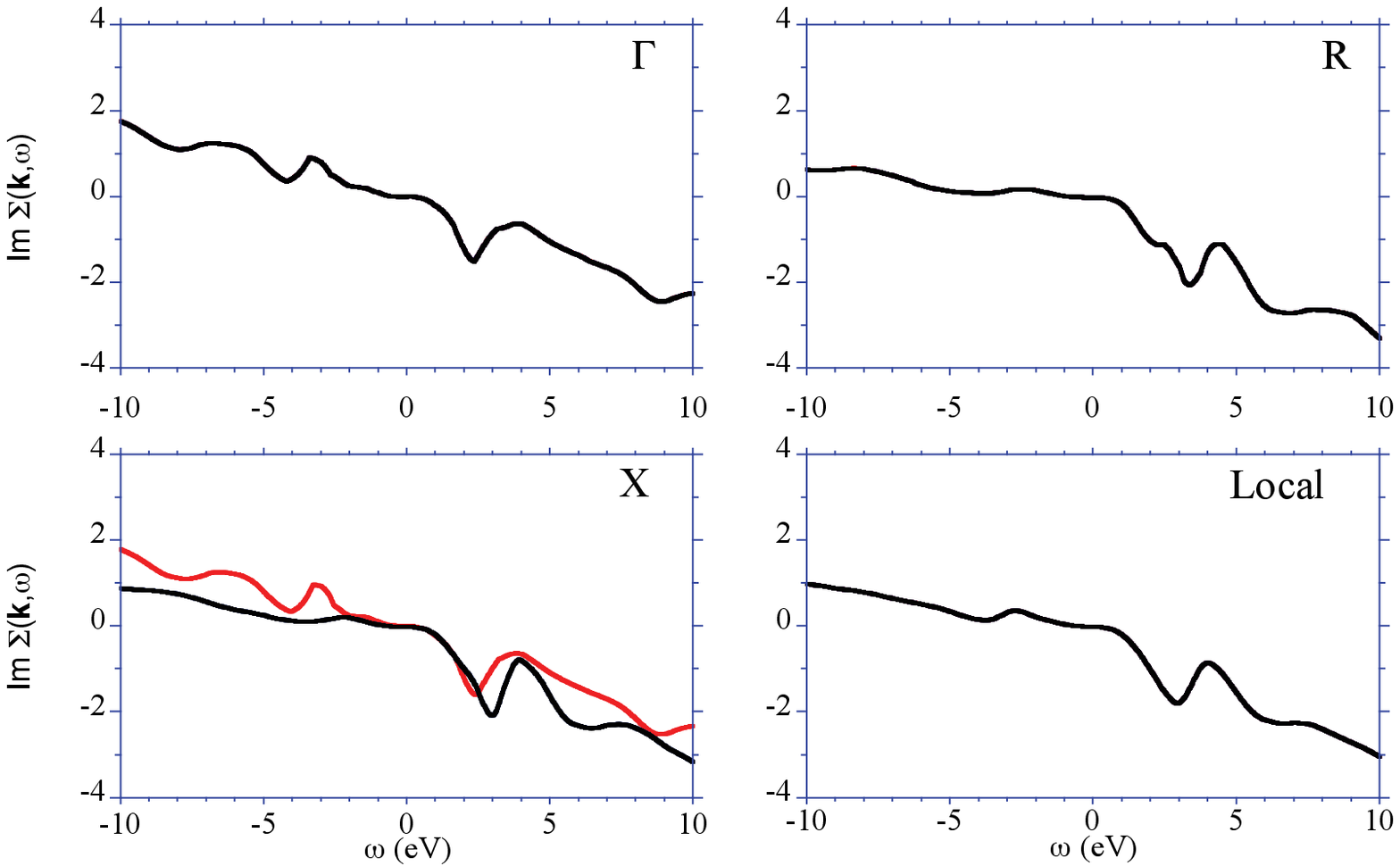}
\end{center}
\caption{(Color online) Self-energy at $\Gamma$, X and R points. The local
self-energy is also shown. Upper panel shows the real part of the self-energy.
The straight line represents $y=\omega- \epsilon_{\mathbf{k}n}$, where
$\epsilon_{\mathbf{k}n}$ is the LDA eigenvalue. Lower panel is the imaginary
part. }%
\label{fig:sig}%
\end{figure}

\begin{figure}[ptb]
\begin{center}
\includegraphics[width=80mm]{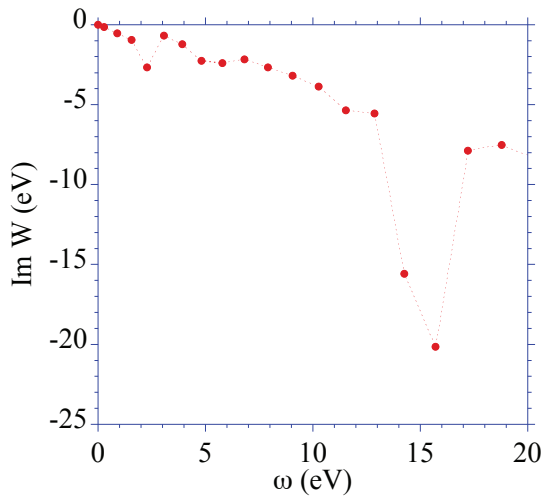}
\end{center}
\caption{(Color online) Imaginary part of the screened Coulomb interaction in
the Wannier basis. }%
\label{fig:imw}%
\end{figure}

We first examine the results for the real and imaginary part of the
self-energy at some representative $\mathbf{k}$-points $\Gamma,X,$ and $R$
shown in Fig.\ref{fig:sig}. $\operatorname{Im}\Sigma$ of the occupied states
are distinctly different from those of the unoccupied states. The former have
a peak structure below the Fermi level at around $-3$ eV while such structure
is essentially absent for the latter. The structure of $\operatorname{Im}%
\Sigma$ can be understood by taking the matrix element $\operatorname{Im}%
\Sigma$ in Eq. (\ref{Gamma1}) and (\ref{Gamma2}):%

\begin{align}
\Gamma_{m}(\mathbf{q},\omega &  \leq\mu)=\sum_{\mathbf{k}n}^{\text{occ}%
}\left\langle \psi_{\mathbf{q}m}\psi_{\mathbf{k}n}|B(\varepsilon_{\mathbf{k}%
n}-\omega)|\psi_{\mathbf{k}n}\psi_{\mathbf{q}m}\right\rangle \nonumber\\
&  \ \ \ \ \ \ \ \ \ \ \ \ \ \ \ \ \ \ \times\theta(\varepsilon_{\mathbf{k}%
n}-\omega),\label{Gammam1}\\
\Gamma_{m}(\mathbf{q},\omega &  >\mu)=\sum_{\mathbf{k}n}^{\text{unocc}%
}\left\langle \psi_{\mathbf{q}m}\psi_{\mathbf{k}n}|B(\omega-\varepsilon
_{\mathbf{k}n}|\psi_{\mathbf{k}n}\psi_{\mathbf{q}m}\right\rangle \nonumber\\
&  \ \ \ \ \ \ \ \ \ \ \ \ \ \ \ \ \ \ \times\theta(\omega-\varepsilon
_{\mathbf{k}n}). \label{Gammam2}%
\end{align}
Since $B$ is proportional to $\operatorname{Im}W$ the structure in
$\operatorname{Im}\Sigma$ is essentially determined by the structure in
$\operatorname{Im}W$ with intensity governed by the overlap between the state
$\psi_{\mathbf{q}m}$ and the occupied or unoccupied states $\psi_{\mathbf{k}%
n}$. Whether the structure in $\operatorname{Im}W$ is carried over to
$\operatorname{Im}\Sigma$ depends on the character of the state $\psi
_{\mathbf{q}m}$. If the state $\psi_{\mathbf{q}m}$ is occupied, there will be
a strong overlap with occupied states $\psi_{\mathbf{k}n}$ so that the
intensity of $\Gamma(\omega\leq\mu)$ may be expected to be stronger than the
intensity of $\Gamma(\omega>\mu).$ Indeed, the peak structure in
$\operatorname{Im}\Sigma$ for the unoccupied state at the $X$ and $R$ point is
stronger above the Fermi level than below. This structure can be traced back
to the peak structure in $\operatorname{Im}W$ at around $1-2$ eV as can be
seen in Fig \ref{fig:imw}.

$\operatorname{Re}\Sigma$ of the occupied states have more structure around
$-3$ eV than for the unoccpied states. This structure originates from the peak
structure in $\operatorname{Im}\Sigma$ at around the same energy. The straight
lines represent%

\[
y=\omega-\varepsilon_{\mathbf{k}n}%
\]
whose intersection with or proximity to the real part of the self-energy below
or above the quasiparticle energy signals the formation of satellites,
provided the imaginary part of the self-energy at the intersection energy is
sufficiently small to produce a discernible feature. It is clear from the
figures that the straight lines only cross $\operatorname{Re}\Sigma$ at one point 
so that a well-defined satellite feature is not expected. However, for the unoccupied
states at $X$ and $R$ the straight lines come close to the proximity of
$\operatorname{Re}\Sigma$ and we expect a formation of a weak satellite at
around $4$ eV above the Fermi level.

To see the difference between the local self-energy and the $\mathbf{k}$-dependent
self-energy we plot in Fig. \ref{fig:sig} the real part of the self-energy at
some $\mathbf{k}$-points and compare them with the local self-energy, which is
the average of the self-energy over the Brillouin zone as defined in Eq.
(\ref{Siglocmm}). The imaginary part of the local self-energy is also shown.
It resembles the self-energy of the unoccupied states because the sum in Eq.
(\ref{Sigmalocal}) is dominated by the unoccupied states since the number of
unoccupied states is much larger than that of the occupied states. The
contribution from the small number of occupied states corresponding to one
$t_{2g}$ electron is weighted down by the contribution from the rest of the
$\mathbf{k}$-points. The real part of the local self-energy is shown in
Fig.\ref{fig:sig} and similar to the imaginary part, it resembles that of the
unoccupied states with little structure below the Fermi level. A\ local $GW$
self-energy would therefore not produce any distinct satellite below the Fermi
level at around $-3$ eV.

\section{Quasiparticle band structures}

\begin{figure}[ptb]
\begin{center}
\includegraphics[width=80mm]{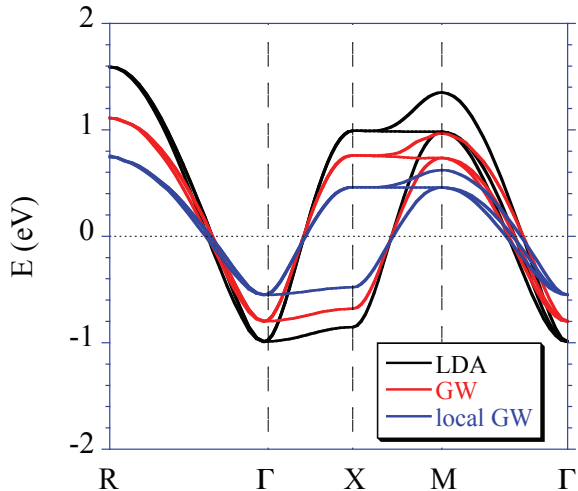}
\end{center}
\caption{(Color online) Quasiparticle bandstructure in LDA, GW, and local GW
approximation. }%
\label{fig:qp}%
\end{figure}

In Fig. \ref{fig:qp} the quasiparticle band structures obtained within the
LDA, the GWA and the local GWA are compared. The effect of the $\mathbf{k}%
$-dependence of the self-energy is striking. The band width obtained from the
full $GW$ calculation is reduced by almost a factor of two when the
$\mathbf{k}$-dependence is neglected. This result is consistent with the
following consideration for the effective mass. The effective mass $m^{\ast}$
is given by%

\begin{equation}
\frac{m}{m^{\ast}}=Z\left[  1+\frac{1}{d\varepsilon/dk}\frac{\partial
\operatorname{Re}\Sigma(k,\omega)}{\partial k}\right]
\end{equation}
where%

\begin{equation}
Z=\left[  1-\frac{\partial\operatorname{Re}\Sigma(k,\omega)}{\partial\omega
}\right]  ^{-1}.
\end{equation}
It is clear from the above expression that if the self-energy is assumed to be
local or onsite then the effective mass or the quasiparticle band width is
determined by the $Z$ factor only since $\operatorname{Re}\Sigma$ is
approximately linear within the band width. However, in the case of SrVO$_{3}%
$, the $GW$ quasiparticle band width is narrowed from the LDA band by only 20
\% whereas the $Z$ factor is about $0.5$. The discrepancy between the band
narrowing and the $Z$ factor can be explained by the $\mathbf{k}$-dependence
of the self-energy. Indeed, when the quasiparticle band structure is
calculated using a local $GW$ self-energy it is found that the band width is
reduced by one half from its LDA value as can be seen in the figure. Thus, the
$\mathbf{k}$-dependent self-energy \emph{widens} the dispersion. This is in
accordance with the electron gas result, in which the free-electron occupied
band width is reduced by only 10 \% by the \emph{GW} self-energy in the
one-shot calculation whereas the $Z$ factor is about $0.7$. In the
self-consistent calculation the free-electron occupied band width is
\emph{widened}, rather than narrowed. This would be contradictory to the fact
that the $Z$ factor is $\sim0.7$ if the band width were determined by the $Z$
factor only. Clearly, the $\mathbf{k}$-dependent self-energy widens the band width.

\begin{figure}[ptb]
\begin{center}
\includegraphics[width=85mm]{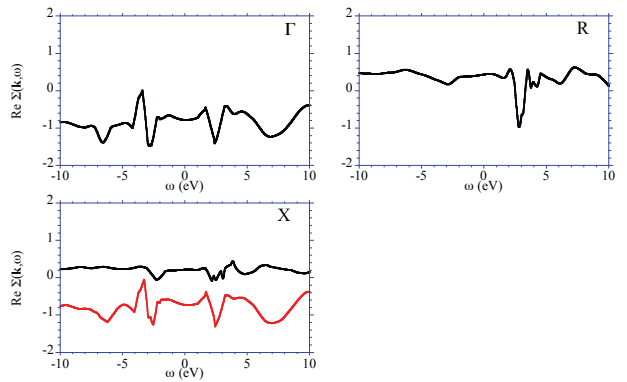}
\end{center}
\caption{(Color online) Real part of the nonlocal self-energy at $\Gamma$, X and R
points. }%
\label{fig:sig2}%
\end{figure}

To understand the origin of the band narrowing when a local self-energy is
used, we plot in Fig. \ref{fig:sig2} the difference between the full and local
self-energies, which represents the nonlocal contribution of the self-energy.
For the occupied states at the $\Gamma$ and $X$ points the difference is
negative whereas for the unoccupied states at the $X$ and $R$ points the
difference is positive. This implies that the occupied states are pushed up
whereas the unoccupied states are pushed down when a local self-energy is
used, resulting in band narrowing. The effect of band narrowing is
particularly revealing for the states at the $X$ point as can be seen in Fig. \ref{fig:sig2}
where the nonlocal self-energies for the occupied and unoccupied states have
different signs.

Although in the present work we have only presented the results for SrVO$_{3}$
we have performed similar calculations on other materials such as iron and
nickel with very similar results. This gives us sufficient confidence to
believe that the results presented in this paper are quite general.

\section{The spectral functions}

\begin{figure}[ptb]
\begin{center}
\includegraphics[width=85mm]{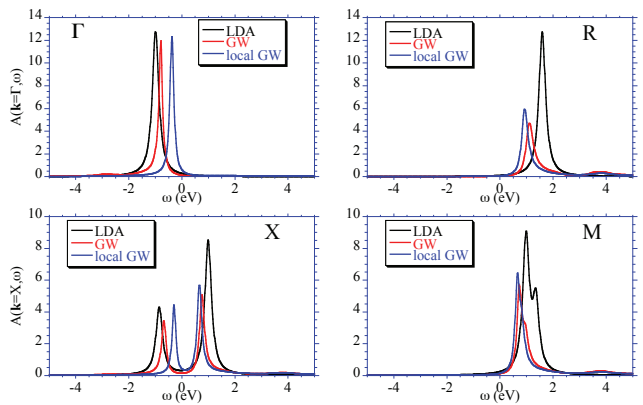}
\end{center}
\caption{(Color online) $\mathbf{k}$-resolved spectral function at $\Gamma$, X, R and M points.
}%
\label{fig:awk}%
\end{figure}

\begin{figure}[ptb]
\begin{center}
\includegraphics[width=80mm]{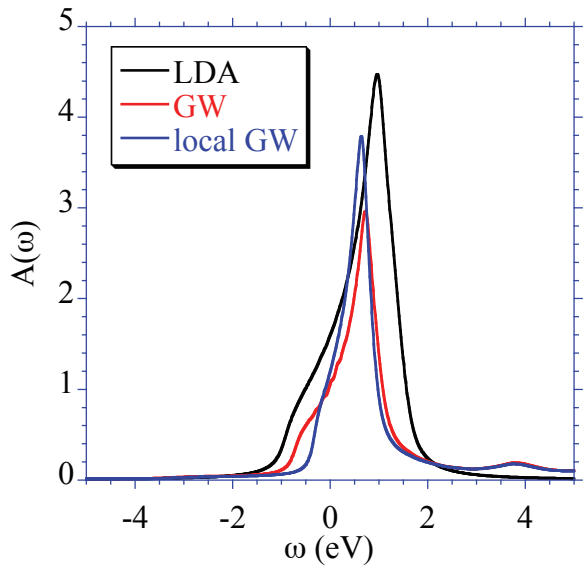}
\end{center}
\caption{(Color online) Total spectral function in LDA, GW and local GW
approximation. }%
\label{fig:aw}%
\end{figure}

In Fig. \ref{fig:awk} we compare the angle-resolved spectral functions at
$\Gamma$, $X$, $R$ and $M$ points obtained from the LDA, the full $GW$
self-energy and the local $GW$ self-energy. As can be already expected from
the plot of $\operatorname{Re}\Sigma$ in Fig. \ref{fig:sig} no strong
satellite structure is expected at the $\Gamma$ point because the straight
line $y=\omega-\varepsilon_{\mathbf{k}n}$ only crosses $\operatorname{Re}%
\Sigma$ at one point corresponding to the quasiparticle energy but away from
this energy it does not come close enough to the proximity of
$\operatorname{Re}\Sigma$. Indeed the (total) spectral function shown in Fig.
\ref{fig:aw} only shows a very weak structure between $-2$ and $-3$ eV with no
satellite feature above the Fermi level. Experimentally, a satellite at $-1.5$
eV is
observed.\cite{morikawa95,inoue98,sekiyama04,yoshida05,takizawa09,yoshida10,yoshimatsu11,aizaki12}%

The spectral functions for the $X$ and $R$ points on the other hand show a
broad but noticeable satellite feature at about $3.5-4.0$ eV above the Fermi
level but with no feature below the Fermi level. This is consistent with the
results for $\operatorname{Re}\Sigma$ shown in Fig. \ref{fig:sig} where the
straight line corresponding to the unoccupied states come close to a peak in
$\operatorname{Re}\Sigma$ around $4$ eV. Since there is only one $d$ electron,
we expect that the satellite above the Fermi level corresponding to
configuration with two electrons is stronger than the one below corresponding
to the removal of the electron.

It is known that the GWA tends to overestimate the position of the satellite
peak arising from a plasmon excitation. The position of the plasmon peak in
$W$ is well described by the RPA but the $GW$ self-energy is first-order in
$W$. It is this first-order approximation that places the plasmon peak
satellite in the spectral function too high in energy. It is reasonable to
expect a similar tendency in low-energy satellites that may be associated with
the Hubbard bands. Indeed, the experimentally observed satellite structures
have lower energies compared with the calculated ones.

\section{Conclusions}

The present work reveals that even in materials with narrow bands, the
nonlocal self-energy has a significant effect on the band structure. The band
width obtained by neglecting the nonlocal self-energy is found to be too
narrow compared to the result of a full calculation and consequently the
effective mass is overestimated. Although the present investigation has been
based on the GWA, it is feasible that the result is general. This result
indicates that calculations based on DMFT using a dynamic $U$ where the
self-energy is local may overestimate the effective mass but further
investigations are needed before a definite conclusion can be reached because
it is not clear how the local self-energy in the present work is related to
the self-energy in DMFT. A promising approach for including the nonlocal
self-energy on top of the DMFT self-energy is the combination of the $GW$
method and DMFT\cite{biermann03} where local correlations responsible for the
formation of Hubbard bands not well described by the GWA is taken care of by
the DMFT while the nonlocal self-energy responsible for modification of the
quasiparticle energy is accounted for by the GWA.

\begin{acknowledgments}
This work was partly supported by HPCI Strategic Programs for Innovative
Research (SPIRE), CMSI, and by KAKENHI (No. 22104010) from MEXT, Japan. FA
acknowledges support from the Swedish Research Council. The calculations were
performed at the supercomputer centers of ISSP, University of Tokyo.
\end{acknowledgments}

\end{document}